\documentclass[12pt]{article}
\usepackage{amssymb,graphicx}
\usepackage{amsmath}
\begin{document}

\title{Improving the estimation of the odds ratio \\using auxiliary information}
\author{{\sc Camelia Goga }$^1$ \quad and \quad {\sc Anne Ruiz-Gazen}$^2$ \\
$^1$Institut de Math\'ematiques de Bourgogne, Universit\'e de Bourgogne,
 Dijon, France \\
$^2$ Toulouse School of Economics, Universit\'e Toulouse 1 Capitole,
Toulouse, France\\
email : camelia.goga@u-bourgogne.fr, anne.ruiz-gazen@tse-fr.eu}

\maketitle

\baselineskip=24pt

\centerline{Abstract}
The odds ratio measure is used in health and social surveys where the odds
of a certain event is to be compared between two populations. It is defined using logistic regression, and requires that data from surveys are accompanied by their weights.
A nonparametric estimation method that incorporates survey weights and auxiliary information may improve the precision of the odds ratio estimator.
It consists in $B$-spline calibration which can handle the nonlinear structure of the parameter. The variance is estimated  through linearization.
Implementation is possible through standard survey softwares. The gain in precision depends on the data as shown on two examples.\\


\noindent {\bf Running title:} Odds ratio estimation in surveys

\section{Introduction }

We study the use of nonparametric weights for estimating the odds ratio when the risk variable, which is the explanatory variable in the logistic regression, is either a continuous or a binary variable.
The odds ratio is used to describe the strength of association
or non-independence between two binary variables defining two groups experiencing a particular event. One binary variable defines a group at risk and a group not at risk; the second binary variable defines the presence or absence of an event related to health.  The odds ratio is the ratio of the odds of the event occurring in one group to
the odds of the same event occurring in the other group. An odds ratio equal to 1 means that the
event has the same odds in both groups; an odds ratio greater than 1 means that the event
has a larger odds in the first group; an odds ratio under 1 means that the event
has a smaller odds in the first group.

When both variables are categorical, the odds ratio estimator is obtained from a contingency table, as the ratio of the estimated row ratios, then, as a function of four numbers. As suggested  by a reviewer, this definition leads to an estimator which takes survey weights into account and yields confidence intervals after linearization. However, this simple definition is not adapted to a continuous risk variable. In this case, the odds ratio measures the change in the odds for an increase of one unit in the risk variable. It is defined through the logistic regression.

For a binary risk variable, the odds ratio is the exponential of the difference of two logits, the logit function being the link function in the logistic regression.
So the logistic regression coefficient for a binary risk variable
corresponds to the logarithm of the odds ratio associated with this risk variable, net the effect of the other variables.
When the risk variable is continuous, the regression coefficient represents the logarithm of the odds ratio associated with a change
in the risk variable of one unit, net the effect of  the other variables. The regression coefficient is a solution of a population estimating equation using the theory developed in Binder (1983) for making inference.
The sampling design  must not be neglected especially for cluster sampling (Lohr, 2010).
Korn and Graubard (1999) and Heeringa et al. (2010) give details and examples of estimating an odds ratio but ignore auxiliary information. Korn and Graubard (1999: 169-170) advocate the use of weighted odds ratios contrary to Eideh and Nathan (2006).  Rao et al. (2002) suggest using post-stratification
information to estimate parameters of interest obtained as solution of an estimating equation. The vector of parameters in the  logistic regression is an example.  Deville (1999) suggested ``allocating a weight $w_k$ to any point in the sample and zero to any other point, regardless of the origin of the weights (Horvitz-Thompson or calibration).'' Goga and Ruiz-Gazen (2014) use auxiliary information to estimate nonlinear parameters through nonparametric weights. The solutions of estimating equations are nonlinear but  Goga and Ruiz-Gazen (2014) give no detail. Our project is the estimation of the odds ratio with auxiliary information.

In  Section 2, we recall the definition
of the odds ratio and express the $B$-spline calibration estimator. In Section 3, we use linearization to derive the asymptotic variance of the estimator under broad assumptions. We infer a variance estimator together with asymptotic normal confidence intervals. In Section 4, we draw guidelines for practical implementation and show the properties of our estimator on two case studies.

\section{Estimation of the odds ratio with survey data}
\subsection{Definition of the parameter}
The odds ratio, denoted by OR, is used to quantify the association between the levels of a response variable
$Y$ and a risk variable $X.$ The value taken by $Y$ is $y_i$ and the value taken by $X$ is $x_i$ for the $i$-th individual in a population $U=\{1, \ldots, N\}.$

The logistic regression
\begin{equation}
\mbox{logit}(p_i)=\log \frac{p_i}{1-p_i}={\beta}_0+{\beta}_1x_i,
\end{equation}
where $p_i=P(Y=1|X=x_i)$ implies that
\begin{equation}
p_i=\exp({\beta}_0+{\beta}_1x_i)(1+\exp({\beta}_0+{\beta}_1x_i))^{-1}.
\end{equation}

The odds ratio is (Agresti, 2002):
\begin{equation}
\mbox{OR}  =  \frac{\mbox{odds}(Y=1|X=x_i+1)}{\mbox{odds}(Y=1|X=x_i)} = \exp{\beta_1}. \label{expbeta}
\end{equation}
With a binary variable $X$, the OR has a simpler form and can be derived from a contingency table. The OR is equal to
\begin{eqnarray}
\mbox{OR}=\frac{N_{00}N_{11}}{N_{01}N_{10}},\label{or_quali}
\end{eqnarray}
where $N_{00},$ $N_{01},$ $N_{10}$, and $N_{11}$ are the population counts associated with the contingency table. In order to estimate the OR of Eq.~(\ref{expbeta}), we estimate first the regression coefficient ${\beta}'=(\beta_0,\beta_1)$ by $\hat{{\beta}}'=(\hat{\beta}_0,\hat{\beta}_1)$, where $\mbox{x}'$ denotes the transpose of $\mbox{x}. $ Eq.~(\ref{expbeta}) yields the estimator of OR:
\begin{equation}
\widehat{\mbox{OR}}=\exp{\hat{\beta}_1}.
\end{equation}
The regression parameters $\beta_0$ and $\beta_1$ are
obtained by maximization of the population likelihood:
\begin{equation}
L(y_1, \ldots, y_N;{\beta})=\prod_{i \in U}\, p_i^{y_i}\, (1-p_i)^{1-y_i}.
\end{equation}
 The maximum likelihood estimator of  ${\beta}$ satisfies:
\begin{eqnarray}
\sum_{i\in U}(y_i-p_i) & = & 0\, , \label{eq:ml1}\\
\sum_{i \in U}(y_i-p_i)x_i & = & \, 0.  \label{eq:ml2}
\end{eqnarray}
Let $\mbox{x}_i=(1 \quad x_i)'$ and $\mu(\mbox{x}'_i{\beta})=\exp(\mbox{x}'_i{\beta})
(1+\exp(\mbox{x}'_i{\beta}))^{-1}. $ We write Eq.~(\ref{eq:ml1}) and (\ref{eq:ml2})  in the equivalent form
\begin{eqnarray}
\sum_{i \in U}\mbox{x}_i(y_i-\mu(\mbox{x}'_i{\beta}))=0\label{maxvrais}
\end{eqnarray}
or, with $\mbox{t}_i({\beta})=\mbox{x}_i(y_i-\mu(\mbox{x}'_i{\beta})),$
\begin{eqnarray}
\sum_{i \in U}\mbox{t}_i({\beta})=0\label{def_t}.
\end{eqnarray}
The regression estimator of ${\beta}$ is defined as an implicit solution of the estimating Eq.~(\ref{maxvrais}). We use iterative methods to compute it.

\subsection{The $ B$-spline nonparametric calibration}\label{b_spline_calibration}

\noindent For  $s$ a sample selected from the population $U$ according to a sample design $p(\cdot)$ , we denote by
$\pi_i >0$ the probability of unit $i$ to be selected in the sample and $\pi_{ij}>0$ the joint  probability of units $i$ and $j$ to be selected in the sample with $\pi_{ii}=\pi_i$. We look for an estimator of ${\beta}$ and of OR  taking the auxiliary variable $Z$, with values $z_1, \ldots, z_N$,  into account.

Deville and S\"arndal (1992) suggest deriving the calibration weights $w_{ks}$  as close as possible to the Horvitz-Thompson sampling weights $d_i=1/\pi_i$  while satisfying the calibration
constraints on known  totals $Z$:
\begin{equation}
\sum_{i \in s} w_{is}z_i  = \sum_{i \in U}z_i.
\end{equation}
This method works well for a linear
relationship between the main and the auxiliary variables. When this relationship is no longer linear,
the calibration constraints must be changed while keeping the property that
the obtained weights do not depend on the main variable.
Basis functions that are more general than the ones defined by  constants and  $z_i$, include $B$-splines,  which are simple to use (Goga and Ruiz-Gazen, 2013), truncated polynomial basis functions, and wavelets.

\subsubsection{$B$-spline functions}\label{sec:bsp}

Spline functions  are used to model nonlinear trends.
A spline function of degree $m$ with $K$ interior knots is a piecewise polynomial
of degree $m-1$ on the intervals between knots, smoothly connected at knots.

\begin{figure}[htbp]
\centerline{\resizebox{0.9\textwidth}{0.5\textheight}
{\includegraphics{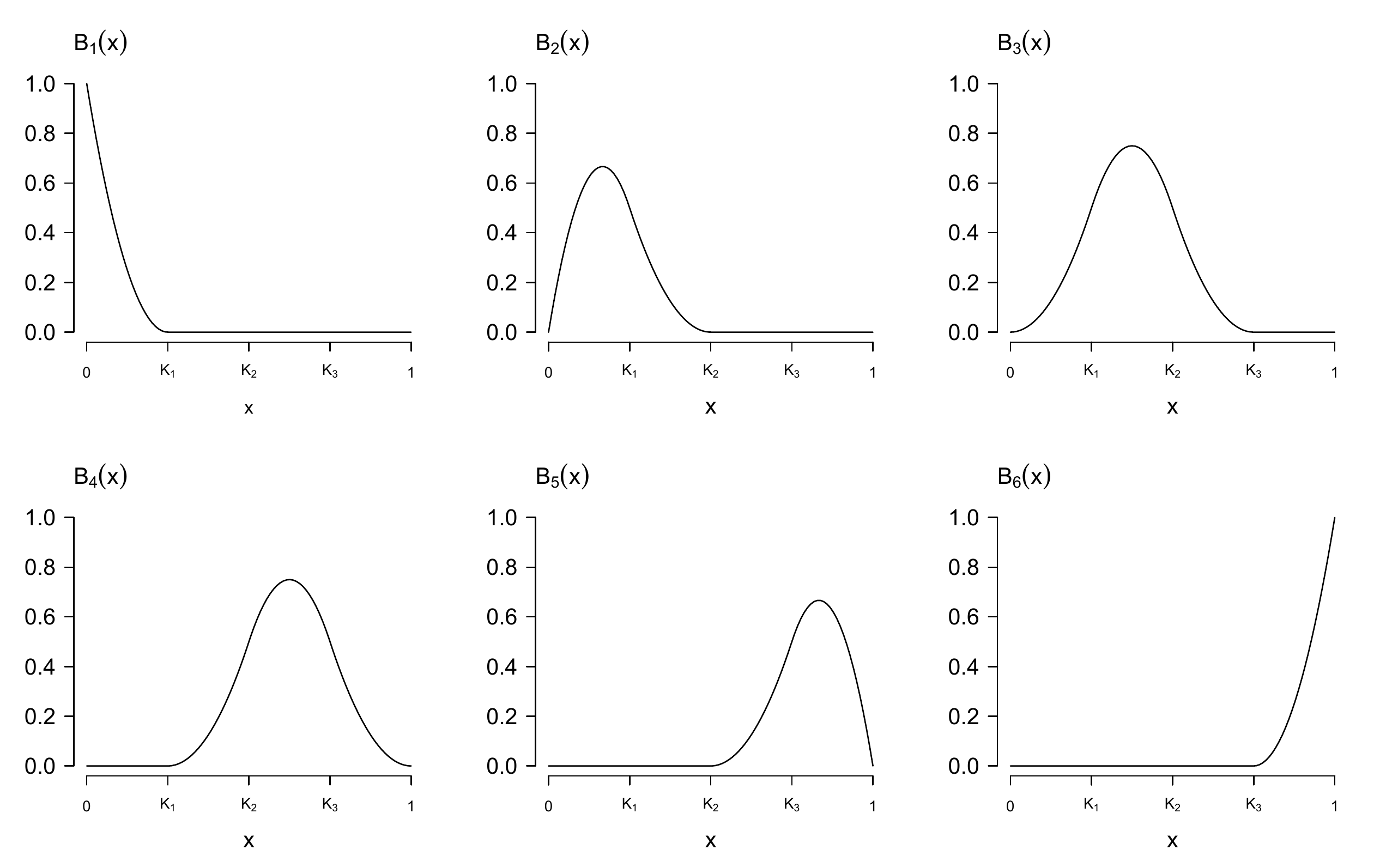}}}
\caption{$B$-spline basis functions with $K=3$ interior knots and $m=3$.}
\label{base_Bspline}
\end{figure}

\begin{figure}[htbp]
\centerline{\resizebox{0.8\textwidth}{0.4\textheight}
{\includegraphics{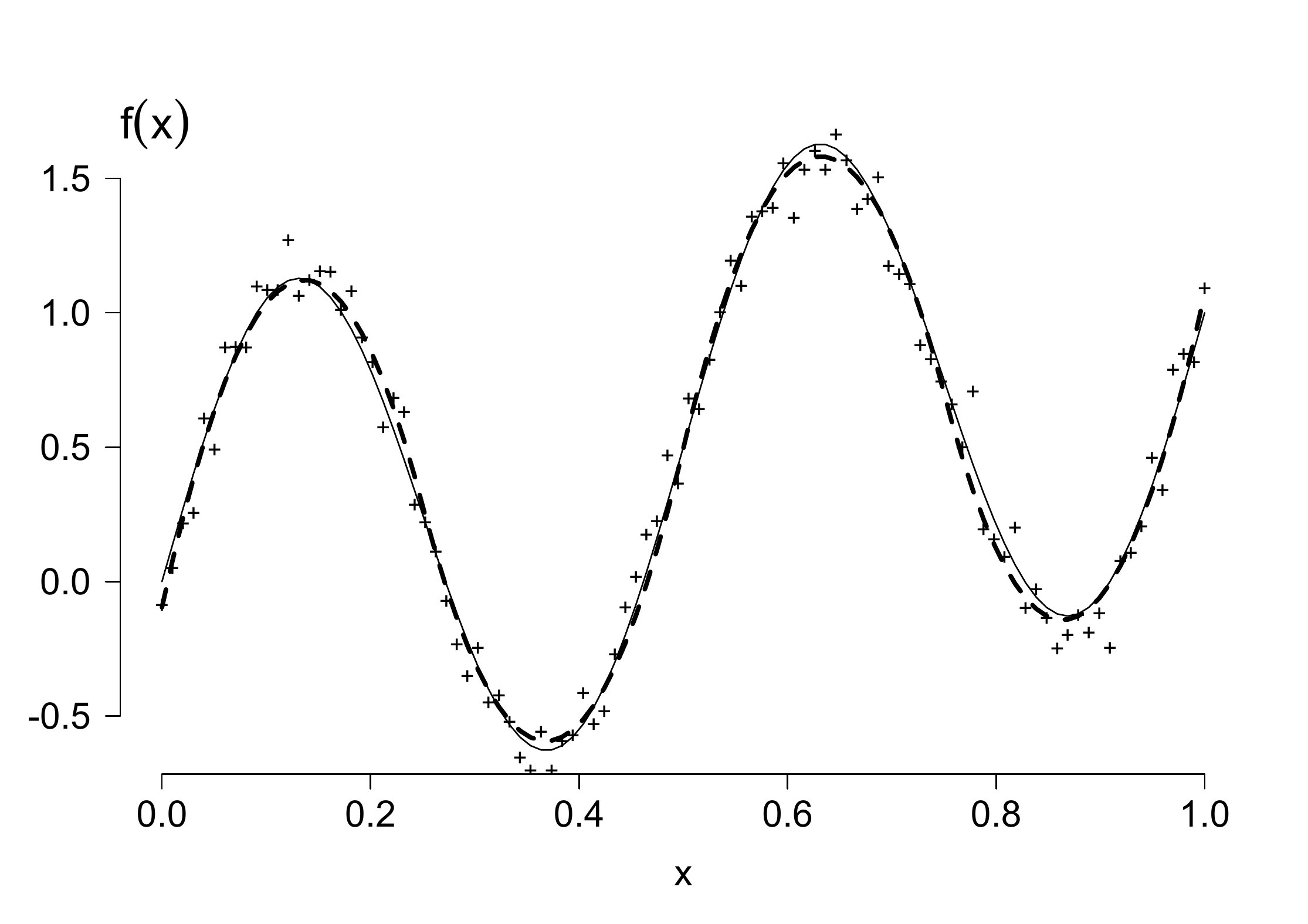}}}
\caption{$B$-spline approximation of $f(x)=x+\sin(4\pi x)$ with $K=3$ interior knots and $m=3$. The crosses correspond to the noisy data.
The solid line is the true function $f$; the dashed line is the $B$-spline approximation. }
\label{approx_sinus}
\end{figure}

The $B$-spline functions $B_1, \ldots, B_q$ of degree $m$ with $K$ interior knots, $q=m+K$ are among the possible basis functions (Dierckx, 1993).
Other basis functions
exist such as the truncated power basis (Ruppert et al., 2003).
For $m=1,$ the $B$-spline basis functions are step functions with jumps at the knots; for $m=2, $
they are piecewise linear polynomials, and so on. Figure \ref{base_Bspline} shows the six $B$-spline basis functions obtained for $m=3$
and $K=3$. Figure \ref{approx_sinus} gives the approximation  of the curve $f(x)=x+\sin(4\pi x)$ taking the noisy data points into account and using the $B$-spline basis.
Even if the function $f$ is nonlinear, the $B$-spline approximation almost coincides with it.
The user chooses
 the spline degree $m$ and the total number $K$ of knots. There is no general rule giving the total number of knots but Ruppert al. (2003) recommend  $m=3$  or $m=4$ and no more than
30 to 40 knots. They also give a simple rule for choosing $K$ (Ruppert et al., 2003: 126).
Usually, the knots are located  at the quantiles of the explanatory variable
(Goga and Ruiz-Gazen, 2013).

\vspace{0.5cm}

\subsubsection{Nonparametric calibration with $B$-spline functions}

The $B$-splines calibration weights $w_{is}^b$ are solution of the optimization problem:
\begin{equation}
(w_{is}^b)_{i\in s} =  \mbox{argmin}_{w}\,\sum_{i \in s}\frac{(w_i-d_i)^2}{q_id_i}
\end{equation}
subject to
\begin{equation}
\sum_{i \in s} w_{is}^b\mbox{b}(z_i) = \sum_{i \in U}\mbox{b}(z_i),
\end{equation}
where  $\mbox{b}(z_i)=(B_1(z_i), \ldots, B_q(z_i))'$  and $q_i$ is a positive constant. They are given by
\begin{eqnarray}
w_{is}^b=d_i\left(1-q_i\mbox{b}'(z_i)(\sum_{i\in s}d_iq_i\mbox{b}(z_i)\mbox{b}'(z_i))^{-1}(\hat t_{b,d}-t_{b})\right)\label{weightcalageB}
\end{eqnarray}
with  $\hat t_{b,d}=\sum_{i\in s}d_i\mbox{b}(z_i), $
$t_{b}=\sum_{i \in U}\mbox{b}(z_i). $ The weights $w_{is}^b$  depend only on the auxiliary variable  and are similar to
Deville and S\"arndal's weights.  The calibration equation implies  $\sum_{i \in s}w_{is}^b=N$ and
$\sum_{i \in s}w_{is}^bz_i=\sum_{i \in U}z_i.$
If $q_i=1$ for all $i\in U,$ we obtain (Goga, 2005):
\begin{eqnarray}
w_{is}^b=d_i t_{b}'\left(\sum_{k\in s}d_k\mbox{b}(z_k)\mbox{b}'(z_k)\right)^{-1}\mbox{b}(z_i).\label{poidsNP}
\end{eqnarray}
Goga and Ruiz-Gazen (2014) use these weights to estimate  totals for variables, which are related
nonlinearly to the auxiliary information and  to estimate nonlinear parameters such as a Gini index.
We use $w_{is}^b$ to estimate the logistic regression coefficient and the odds ratio efficiently.

\subsection{Estimation of OR using $B$-spline nonparametric calibration}\label{section_np}

The regression coefficient ${\beta}$ is a nonlinear finite population function defined by the implicit  Eq.~(\ref{maxvrais}).
The functional method by Deville (1999), specified for the nonparametric case by Goga and Ruiz-Gazen (2014), is used
to build a nonparametric estimator of ${\beta}$ defined through the weights of Eq.~(\ref{poidsNP}).
$M$ is the finite measure assigning the unit mass to each $y_i$, $i\in U$, and zero elsewhere:
\begin{equation}
M=\sum_{i \in U}\delta_{y_i}
\end{equation}
where $\delta_{y_i}$ is the Dirac function at $y_i,$ $\delta_{y_i}(y)=1$ for $y=y_i$ and zero elsewhere.
The functional $T$ defined with respect to the measure $M$ and depending on the parameter ${\beta}$ defined by
\begin{equation}
T(M; {\beta})=\sum_{i \in U}\mbox{x}_i(y_i-\mu(\mbox{x}'_i{\beta})).
\end{equation}
The regression coefficient ${\beta}$ is the solution of the implicit equation
\begin{eqnarray}
T(M; {\beta})=0\label{betafunct}. \label{eq:sco}
\end{eqnarray}
Eq.~(\ref{eq:sco})  is called the score equation.
\noindent The measure $M$ may be estimated using the Horvitz-Thompson weights $d_k=1/\pi_k$ or the linear calibration weights (Deville, 1999). We suggest
 using the nonparametric weights derived in Eq.~(\ref{poidsNP}):
\begin{eqnarray}
w_{is}^b=d_i\left(\sum_{k\in U}\mbox{b}(z_k)\right)'\left(\sum_{k\in s}d_k\mbox{b}(z_k)\mbox{b}'(z_k)\right)^{-1}\mbox{b}(z_i)
\end{eqnarray}
and estimate $M$ by
\begin{eqnarray}
\widehat M=\sum_{i\in s}w^b_{is}\delta_{y_i}.
\end{eqnarray}

\noindent Plugging $\widehat M$ into the functional expression of $ {\beta}$ given by Eq.~(\ref{betafunct}) yields the $B$-spline
calibrated estimator  $\widehat{{\beta}}$ of ${\beta}$:
 \begin{eqnarray}
 T(\widehat M;\widehat{{\beta}})=0, \label{estimbetafunct}
 \end{eqnarray}
 which means that  $\widehat{{\beta}}$ is the solution of the implicit equation:
 \begin{eqnarray}
 \sum_{i \in s}w_{is}^b\mbox{x}_i(y_i-\mu(\mbox{x}'_i\widehat{{\beta}}))=0.\label{maxvraisech}
 \end{eqnarray}
The functional method allows us to incorporate auxiliary information for estimating the logistic regression coefficient and any parameter ${\beta}$ defined
as a solution of estimating equations.

\noindent The functional $T$ is differentiable with respect to ${\beta}$ and
\begin{equation}
\frac{\partial T}{\partial{{\beta}}}=-\sum_{i\in U}\nu(\mbox{x}'_i{\beta})\mbox{x}_i\mbox{x}'_i=\mbox{X}'
{\Lambda({\beta})}\mbox{X}:=\mbox{J}({\beta}), \label{jacobian_pop}
\end{equation}
with $\mbox{X}=(\mbox{x}'_i)_{i\in U}$ and ${\Lambda({\beta})}=-\mbox{diag}(\nu(\mbox{x}'_i{\beta}))$
with $\nu(\mbox{x}'_i{\beta})=\mu(\mbox{x}'_i{\beta})(1-\mu(\mbox{x}'_i{\beta}))$ the derivative of $\mu$. The $2\times 2$ matrix $\mbox{X}'{\Lambda({\beta})}\mbox{X}$ is invertible and  $\mbox{J}({\beta})$
is  definite negative.  From Eq.~(\ref{jacobian_pop}), the matrix  $\mbox{J}({\beta})$ is a total estimated using the nonparametric weights  $w_{is}^b$ by:
\begin{eqnarray}
\widehat{\mbox{J}}_w({\beta})=-\sum_{i \in s}w_{is}^b\nu(\mbox{x}'_i{\beta})\mbox{x}_i\mbox{x}'_i=
\mbox{X}_s'\widehat{{\Lambda}}({\beta})\mbox{X}_s, \label{estimJ}
\end{eqnarray}
 where $\widehat{{\Lambda}}({\beta})=-\mbox{diag}(w_{is}^b\nu(\mbox{x}'_i{\beta}))_{i\in s}$ and
 $\mbox{X}_s=(\mbox{x}'_i)_{i\in s}$.

An iterative Newton-Raphson method is used to compute $\widehat{{\beta}}.$ The $r$-th step of the Newton-Raphson
algorithm is:
\begin{equation}
\widehat{{\beta}}_r=\widehat{{\beta}}_{r-1}-\widehat{\mbox{J}}_w(\widehat{{\beta}}_{r-1})T(\widehat M;
\widehat{{\beta}}_{r-1}),
\end{equation}
where $\widehat{{\beta}}_{r-1}$ is the value of $\widehat{{\beta}}$ obtained at the $(r-1)$-th step.
$\widehat{\mbox{J}}_w(\widehat{{\beta}}_{r-1})$ is the value of
$\widehat{\mbox{J}}_w({\beta})$ and $T(\widehat M; \widehat{{\beta}}_{r-1})$ the value of $T(\widehat M;{\beta})$ evaluated at
${\beta}=\widehat{{\beta}}_{r-1}. $ Iterating to convergence produces the nonparametric estimator
$\widehat{{\beta}}$ and the estimated Jacobian matrix $\widehat{\mbox{J}}_w(\widehat{{\beta}}).$
The odds ratio is estimated  by $\widehat{\mbox{OR}}=\exp(\hat{\beta_1})$ and $\widehat{\mbox{J}}_w(\widehat{{\beta}})$
is used in section \ref{sec:var} to estimate the variance of $\hat{{\beta}}.$

\section{Variance estimation and confidence intervals}\label{sec:var}
\subsection{Variance estimation}
The coefficient ${\beta}$ of the  logistic regression is nonlinear and nonparametric weights $w^b_{is}$ to estimate ${\beta}$ add more nonlinearity.   We approximate $\widehat{{\beta}}$ in Eq.~(\ref{estimbetafunct}) by a linear estimator in two steps: we  first treat the nonlinearity due to  ${\beta}$, and second the nonlinearity due to the nonparametric estimation.
This procedure is different  from Deville (1999).

From the implicit function theorem, there exists a unique functional $\widetilde T$ such that
\begin{equation}
\widetilde T(M)={\beta}\;\;\;\mbox{ and } \;\;\;
\widetilde T(\widehat M)=\widehat{{\beta}}.
\end{equation}
Moreover, the functional $\widetilde T$ is also Fr\'echet differentiable with respect to $M$. The derivative of $\widetilde T$
with respect to $M$, called the influence function, is defined by
\begin{eqnarray} \label{if}
I\widetilde T(M,\xi)=\lim_{\lambda\rightarrow 0}\,\frac{\widetilde T(M+\lambda \delta_{\xi})-\widetilde T(M)}{\lambda},
\end{eqnarray}
\noindent where $\delta_{\xi}$ is the Dirac function at $\xi . $  We give a first-order expansion  of $\tilde T$ in $\widehat M/N$ around $M/N,$
\begin{eqnarray}
\widetilde T\left(\frac{\widehat M}{N}\right)=\widetilde T\left(\frac{M}{N}\right)+\int_{-\infty}^{+\infty} I\widetilde T\left(\frac{M}{N},\xi\right)d
\left(\frac{\widehat M}{N}-\frac{M}{N}\right)(\xi)+o_p(n^{-1/2}),\label{vonmises1}
\end{eqnarray}
which is also:
\begin{eqnarray}
\widetilde T(\widehat M)=\widetilde T(M)+\int_{-\infty}^{+\infty} I\widetilde T\left(M,\xi\right)d(\widehat M-M)(\xi)+o_p(n^{-1/2}), \label{vonmises2}
\end{eqnarray}
because $\widetilde T$  is a functional of degree zero, namely $\widetilde T(M/N)=\widetilde T(M)$ and $I\widetilde T
 \left(M/N,\xi\right)=N I\widetilde T\left(M,\xi\right)$ (Deville, 1999).

For all $i\in U$, the linearized variable $\mbox{u}_i$   of $\widetilde T(M)={\beta}$ is defined as the value
of the influence function  $I\widetilde T$ at $\xi=y_i$:
\begin{eqnarray}
\mbox{u}_i & = & I\widetilde T(M,y_i)=-\left(\frac{\partial T}{\partial{{\beta}}}\right)^{-1} IT(M,y_i;
{\beta})\nonumber\\
 & = & -\left(\mbox{X}'{\Lambda({\beta})}\mbox{X}\right)^{-1}\mbox{x}_i(y_i-\mu(\mbox{x}'_i{\beta}))
 =-\mbox{J}^{-1}({\beta})\cdot\mbox{x}_i(y_i-\mu(\mbox{x}'_i{\beta})).\label{eq_lin}
\end{eqnarray}
The linearized variable $\mbox{u}_i=(u_{i,0},u_{i,1})'$ is a two-dimensional vector depending on the unknown parameter ${\beta}$ and on totals contained in the matrix $\mbox{J}({\beta})$.
\noindent Eq.~(\ref{vonmises2}) becomes:
\begin{eqnarray}
\hat {{\beta}}-{\beta}\simeq \sum_{i\in s}w_{is}^b\mbox{u}_i-\sum_{i\in U}\mbox{u}_i.\label{linear1}
\end{eqnarray}
The second component $u_{i,1}$  of $\mbox{u}_i,$ is the linearized variable of $\beta_1$. With binary data, the odds ratio is given by Eq.~(\ref{or_quali}), which implies that
\begin{equation}
\ln(\mbox{OR})=\ln(N_{00})+\ln(N_{11})-\ln(N_{01})-\ln(N_{10}).
\end{equation}
In this case, the linearized variable of $\beta_1$ has the expression:
\begin{eqnarray}
u_{i,1}=\frac{\mbox{1}_{\{x_i=0,y_i=0\}}}{N_{00}}+\frac{\mbox{1}_{\{x_i=1,y_i=1\}}}{N_{11}}-\frac{\mbox{1}_{\{x_i=1,y_i=0\}}}{N_{10}}-\frac{\mbox{1}_{\{x_i=0,y_i=1\}}}{N_{01}}
\label{lin1}
\end{eqnarray}
and the same expression is obtained from Eq.~(\ref{eq_lin}) after some algebra.
When the weights $w_{is}^b$ are equal to the sampling  weights, namely $w_{is}^b=1/\pi_i$, Eq.~(\ref{linear1}) implies that the asymptotic variance
of $\hat {{\beta}}$ is:
\begin{eqnarray}
\mbox{AV}(\hat {{\beta}})=\mbox{Var}\left(\sum_{i\in s}d_i\mbox{u}_i\right)=\mbox{J}^{-1}({\beta})\,
\mbox{\mbox{V}}_{\mbox{\sc ht}}(\hat{\mbox{t}}_d({\beta}))\, \mbox{J}^{-1}
({\beta}), \label{var_sans_inf}
\end{eqnarray}
where $\displaystyle \mbox{\mbox{V}}_{\mbox{\sc ht}}(\hat{\mbox{t}}_d({\beta}))$ is the Horvitz-Thompson variance of
$\hat{\mbox{t}}_d({\beta})=\sum_{i\in s} \mbox{t}_i({\beta}) / \pi_i$ with $\mbox{t}_i({\beta})=
\mbox{x}_i(y_i-\mu(\mbox{x}'_i{\beta}))$:
\begin{equation}
\mbox{\mbox{V}}_{\mbox{\sc ht}}(\hat{\mbox{t}}_d({\beta}))=
\mbox{Var}\left(\sum_{i\in s}\frac{\mbox{t}_i({\beta})}{\pi_i}\right)=
\sum_{i\in U}\sum_{i\in U}(\pi_{ij}-\pi_i\pi_j)\frac{\mbox{t}_i({\beta})}{\pi_i}\frac{\mbox{t}_j({\beta})}{\pi_j}.
\end{equation}
 Binder (1983) gives the same  asymptotic expression for the variance.

For $B$-spline basis functions  formed by step functions on intervals between knots ($m=1$), the weights $w_{is}^b$ yield the post-stratified estimator of ${\beta}$
(Rao et al., 2002). Linear calibration weights lead to the case treated by Deville (1999).

For the general case of nonparametric calibration weights $w_{is}^b$, a supplementary linearization step is necessary.
The right hand side of Eq.~(\ref{linear1}) is  a nonparametric calibration estimator for the total of the linearized variable $\mbox{u}_i$. It can be written
as a generalized regression estimator (GREG):
\begin{eqnarray}
\sum_{i\in s}w_{is}^b\mbox{u}_i-\sum_{i\in U}\mbox{u}_i= \sum_{i\in s}\frac{\mbox{u}_i-\widehat{{\theta}}_{u}'\mbox{b}(z_i)}{\pi_i}+\sum_{i\in U}\widehat{{\theta}}_{u}'\mbox{b}(z_i)-
\sum_{i\in U}\mbox{u}_i, \label{greg_var_lin}
\end{eqnarray}
where $\widehat{{\theta}}_{u}=(\sum_{i\in s}d_i\mbox{b}(z_i)\mbox{b}'(z_i))^{-1}(\sum_{i\in s}d_i\mbox{b}(z_i)\mbox{u}'_i). $
We explain the linearized variable by means of a piecewise polynomial  function. This fitting allows more flexibility  and implies that the residuals $\mbox{u}_i-\widehat{{\theta}}_{u}'\mbox{b}(z_i)$
have a smaller dispersion   than with a
linear fitting regression.

 In order to derive the asymptotic variance of the nonparametric calibrated estimator, we assume
 that $||\mbox{x}_i||< C$ for all $i\in U$ with $C$ a positive constant independent of $i$ and $N$.
 The Euclidian norm is denoted $||\cdot||$.  The matrix norm $||\cdot||_2$ is defined by $||\mbox{A}||_2^2=\mbox{tr}(\mbox{A}'\mbox{A}). $ The linearized variable verifies $N||\mbox{u}_i||=O(1)$ uniformly in $i,$  because
\begin{equation}
   N||\mbox{u}_i||\leq ||N\mbox{J}^{-1}({\beta})||_2\,  ||\mbox{x}_i||\, |y_i-\mu(\mbox{x}_i'{\beta}))|=O(1),
\end{equation}
where the Jacobian matrix $\mbox{J}({\beta}) $ contains totals
\begin{equation}
\mbox{J}({\beta})=-\left(\begin{array}{cc}
\sum_{i\in U}\nu(\mbox{x}_i'{\beta}) & \sum_{i\in U}x_i\nu(\mbox{x}_i'{\beta})\\
\sum_{i\in U}x_i\nu(\mbox{x}_i'{\beta}) & \sum_{i\in U}x^2_i\nu(\mbox{x}_i'{\beta})\end{array}\right)
\end{equation}
and
\begin{equation}
\left(\frac{1}{N}\sum_{i\in U}\nu(\mbox{x}_i'{\beta})\right)^2\leq \frac{1}{N}\sum_{i\in U}(\nu(\mbox{x}_i'{\beta}))^2=O(1)
\end{equation} because $\nu(\mbox{x}_i'{\beta})<1.$
Under the assumptions of theorem 7 in Goga and Ruiz-Gazen (2014), the
nonparametric calibrated estimator $\sum_{i\in s}w_{is}^b\mbox{u}_i$ is  asymptotically equivalent to
\begin{eqnarray}
\sum_{i\in s}w_{is}^b\mbox{u}_i-\sum_{i\in U}\mbox{u}_i\simeq \sum_{i\in s}\frac{\mbox{u}_i-\widetilde{{\theta}}_{u}'\mbox{b}(z_i)}{\pi_i}+\sum_{i\in U}\widetilde{{\theta}}_{u}'\mbox{b}(z_i)-
\sum_{i\in U}\mbox{u}_i, \label{linear2}
\end{eqnarray}
where $\widetilde{{\theta}}_{u}=(\sum_{i\in U}\mbox{b}(z_i)\mbox{b}'(z_i))^{-1}\sum_{i\in U}\mbox{b}(z_i)\mbox{u}'_i.$
The variance of $\hat {{\beta}}$ is approximated by the Horvitz-Thompson variance of the residuals
$\mbox{u}_i-\widetilde{{\theta}}_{u}'\mbox{b}(z_i),$
\begin{eqnarray}
\mbox{AV}(\hat {{\beta}})=\mbox{Var}\left(\sum_{i\in s}\frac{\mbox{u}_i-\widetilde{{\theta}}_{u}'\mbox{b}(z_i)}{\pi_i}\right)=\sum_{i\in U}\sum_{i\in U}(\pi_{ij}-\pi_i\pi_j)\frac{\mbox{u}_i-\widetilde{{\theta}}_{u}'\mbox{b}(z_i)}{\pi_i}\frac{\mbox{u}_j-\widetilde{{\theta}}_{u}'\mbox{b}(z_j)}{\pi_j}.\label{var_asymt}
\end{eqnarray}
Eq.~(\ref{linear2}) states that the $B$-spline nonparametric calibration estimator of $\sum_{i\in U}\mbox{u}_i$ is asymptotically equivalent
to the generalized difference estimator. We interpret this result as fitting a nonparametric model on the linearized variable $\mbox{u}_i$
taking into account the auxiliary information $z_i$. Nonparametric models are a good choice when the linearized variable obtained from the first linearization step
does not depend linearly on $z_i$, as it is the case in the logistic regression, which  implies a second linearization step.

We write the asymptotic variance in Eq.~(\ref{var_asymt}) in a matrix form similar to Eq.~(\ref{var_sans_inf}). Consider again
$\mbox{t}_i({\beta})=\mbox{x}_i(y_i-\mu(\mbox{x}'_i{\beta}))$ and let $\widetilde{{\theta}}_{\mathbf t}=(\sum_{i\in s}\mbox{b}(z_i)\mbox{b}'(z_i))^{-1}\sum_{i\in s}\mbox{b}(z_i)\mbox{t}'_i({\beta}).$ We have
\begin{equation}
\mbox{u}_i-\widetilde{{\theta}}_{u}'\mbox{b}(z_i)=-\mbox{J}^{-1}({\beta})\left(\mbox{t}_i({\beta})-
\widetilde{{\theta}}'_{\mathbf t}\mbox{b}(z_i)\right),
\end{equation}
and the asymptotic variance of $\hat {{\beta}}$ is:
\begin{equation}
\mbox{AV}(\hat {{\beta}}) =  \mbox{J}^{-1}({\beta})\, \mbox{V}_{\mbox{\sc ht}}(\hat{{e}}_d({\beta}))\,
\mbox{J}^{-1}({\beta})\label{varasymbeta}
\end{equation}
where $\hat{{e}}_d({\beta})=\displaystyle\sum_{i\in s}\frac{\mbox{e}_i({\beta})}{\pi_i}$ is the
Horvitz-Thompson estimator of the residual  ${e}_i({\beta})=\mbox{t}_i({\beta})-
\widetilde{{\theta}}'_{\mathbf t}\mbox{b}(z_i)$  of $\mbox{t}_i({\beta})$ using $B$-spline calibration.
Eq.~(\ref{varasymbeta}) shows that improving the estimation of ${\beta}$ is equivalent to improving the estimation
of the score equation   $\mbox{t}_i=\mbox{x}_i(y_i-\mu(\mbox{x}'_i{\beta})). $\\
The quantity of interest is the asymptotic variance of $\hat{\beta_1}$. It is the $(2,2)$ element of the matrix $\mbox{AV}(\hat{{\beta}})$ given by 
Eq.~(\ref{var_asymt}). We have $\mbox{u}_i=(u_{i,0},u_{i,1})'$ and
\begin{eqnarray}
\mbox{u}_i-\widetilde{{\theta}}'_{u}\mbox{b}(z_i)=\left(\begin{array}{c}
u_{i,0}-\widetilde{{\theta}}'_{u_0}\mbox{b}(z_i)\\
u_{i,1}-\widetilde{{\theta}}'_{u_1}\mbox{b}(z_i)\end{array}
\right)
\end{eqnarray}
where  $\widetilde{{\theta}}_{u_0}=(\sum_{i\in U}\mbox{b}(z_i)\mbox{b}'(z_i))^{-1}\sum_{i\in U}\mbox{b}(z_i)u_{i,0}$ and $\widetilde{{\theta}}_{u_1}=(\sum_{i\in U}\mbox{b}(z_i)\mbox{b}'(z_i))^{-1}\sum_{i\in U}\mbox{b}(z_i)u_{i,1}.$
We obtain
\begin{equation}
\mbox{AV}(\hat{\beta}_1)=\mbox{Var}\left(\sum_{i\in s}\frac{u_{i,1}-\widetilde{{\theta}}_{u_1}'\mbox{b}(z_i)}{\pi_i}\right).
\end{equation}

\noindent The linearized variable $\mbox{u}_i$ is unknown and is estimated by:
\begin{eqnarray}
\hat{\mbox{u}}_i & = &-\widehat{\mbox{J}}_w^{-1}(\widehat{{\beta}})\, \mbox{x}_i(y_i-\mu(\mbox{x}'_i
\widehat{{\beta}})) \label{varlin_estimator}\\
& = & -\widehat{\mbox{J}}_w^{-1}(\widehat{{\beta}})\, \hat{\mbox{t}}_i
\end{eqnarray}
where the matrix $\widehat{\mbox{J}}_w$ is computed according to Eq.~(\ref{estimJ}) and $\hat{\mbox{t}}_i$ is the estimation of $\mbox{t}_i({\beta})$ for ${\beta}=\widehat{{\beta}}$.
The asymptotic variance $\mbox{AV}(\widehat {{\beta}})$ given in Eq.~(\ref{var_asymt}) is estimated by the Horvitz-Thompson variance estimator
  with $\mbox{u}_i$ replaced by $\hat{\mbox{u}}_i$ given in Eq.~(\ref{varlin_estimator}):
\begin{eqnarray}
\widehat{V}(\widehat {{\beta}}) & = & \widehat{V}_{\mbox{\sc ht}}\left(\sum_{i\in s}\frac{\hat{\mbox{u}}_i-
\widehat{{\theta}}'_{\widehat{\mbox{u}}}\mbox{b}(z_i)}{\pi_i}\right)\label{estimator_variance}
\end{eqnarray}
where $\widehat{{\theta}}_{\widehat{\mbox{u}}}=(\sum_{i\in s}d_i\mbox{b}(z_i)\mbox{b}'(z_i))^{-1}\sum_{i\in s}d_i\mbox{b}(z_i)\hat{\mbox{u}}'_i. $ The variance estimator of $\hat{\beta}_1$ is given by
\begin{eqnarray}
\hat{V}(\hat{\beta}_1)=\mbox{Var}\left(\sum_{i\in s}\frac{\hat{u}_{i,1}-\widehat{{\theta}}_{\hat{u}_1}'\mbox{b}(z_i)}{\pi_i}\right).
\end{eqnarray}
The variance estimator given in Eq.~(\ref{estimator_variance}) can be written in a matrix form. Let $\widehat{{\theta}}_{\widehat{\mathbf t}}=(\sum_{i\in s}d_i\mbox{b}(z_i)\mbox{b}'(z_i))^{-1}\sum_{i\in s}d_i\mbox{b}(z_i)\hat{\mbox{t}}'_i$ and $ \widehat{V}(\widehat {{\beta}})$ is written as:
\begin{eqnarray}
 \widehat{V}(\widehat {{\beta}}) &= & \widehat{\mbox{J}}_w^{-1}(\widehat{{\beta}})\, \widehat{\mbox{V}}_{\mbox{\sc ht}}
 (\hat{{e}}_d(\widehat{{\beta}}))\, \widehat{\mbox{J}}_w^{-1}(\widehat{{\beta}})\label{variance_estim}
\end{eqnarray}
where $\widehat{\mbox{V}}_{\mbox{\sc ht}}(\hat{{e}}_d)$
is the Horvitz-Thompson variance estimator of $\hat{{e}}_d(\hat{{\beta}})$ obtained by replacing
${e}_i({\beta})$ with $\hat{{e}}_i(\hat{{\beta}})=\hat{\mbox{t}}_i-
\widehat{{\theta}}'_{\hat{\mathbf t}}\mbox{b}(z_i), $
\begin{eqnarray}
\widehat{\mbox{V}}_{\mbox{\sc ht}}(\hat{{e}}(\hat{{\beta}}))=\sum_{i\in s}\sum_{i\in s}\frac{\pi_{ij}-\pi_i\pi_j}{\pi_{ij}}
\frac{\hat{{e}}_i(\hat{{\beta}})}{\pi_i}\frac{\hat{{e}}_j(\hat{{\beta}})}{\pi_j}.
\end{eqnarray}

\subsection{Confidence interval for the odds ratio}

The variance estimator of $\hat{\beta}_1$ is obtained from Eq.~(\ref{variance_estim}) as:
\begin{equation}
\hat{V}(\hat{\beta}_1)=\widehat{\mbox{J}}_w^{-1}(\widehat{{\beta}})\, \mbox{V}_{\mbox{\sc ht}}(\hat{e}_2(\widehat{{\beta}}))
\, \widehat{\mbox{J}}_w^{-1}(\widehat{{\beta}}),
\end{equation}
where $\hat{e}_2(\widehat{{\beta}})$ is the second component of  $\hat{{e}}(\widehat{{\beta}})$ so that,
under regularity conditions, the $(1-\alpha)\%$ normal interval for $\beta_1$ is:
\begin{equation}
\mbox{CI}_{1-\alpha}({\beta_1})=\left[\hat{\beta}_1-z_{\alpha/2}\left(\hat{V}(\hat{\beta}_1)\right)^{1/2},
\hat{\beta}_1+z_{\alpha/2}\left(\hat{V}(\hat{\beta}_1)\right)^{1/2}\right],
\end{equation}
where $z_{\alpha/2}$ is the upper $\alpha/2$-quantile of a $\mathcal{N}(0,1)$ variable. Then the confidence interval for OR is:
\begin{equation}
\mbox{CI}_{1-\alpha}(\mbox{OR})=\left[\exp{\left(\hat{\beta}_1-z_{\alpha/2}\left({\hat{V}(\hat{\beta}_1)}\right)^{1/2}\right)}, \exp{\left(\hat{\beta}_1+z_{\alpha/2}
\left({\hat{V}(\hat{\beta}_1)}\right)^{1/2}\right)}\right],
\end{equation}
which is not symmetric around the estimated odds ratio but provides more accurate coverage rates of the true population value for a specified
$\alpha$ (Heeringa et al., 2010).

\section{Implementation and case studies}

\subsection{Implementation}

\begin{enumerate}
\item Compute the $B$-spline basis functions $B_j,$ for $j=1, \ldots, q. $ The $B$-spline basis functions are obtained using \textrm{SAS} or \textrm{R}. The user has only to specify the degree $m$ and the total number of knots.
\item Use the sampling weights $d_i=1/\pi_i$ and the $B$-spline functions to derive the nonparametric weights $w_{is}^b$ and the estimated
parameter ${\beta}$.
\item Compute the linearized variable $\mbox{u}_i$  estimated by $\hat{\mbox{u}}_i.$
\item Compute the estimated predictions $\widehat{{\theta}}_{\widehat{\mbox{u}}}'\mbox{b}(z_i)$ with
\begin{equation}
\widehat{{\theta}}_{\widehat{\mbox{u}}}=\left(\sum_{i\in s}d_i\mbox{b}(z_i)\mbox{b}'(z_i)\right)^{-1}\left(\sum_{i\in s}d_i\mbox{b}(z_i)
\widehat{\mbox{u}}'_i\right)
\end{equation}and the associated residuals $\hat{\mbox{u}}_i-
\widehat{{\theta}}'_{\widehat{\mbox{u}}}\mbox{b}(z_i)$.

\item Use a standard computer software able to compute variance estimators and apply it to the  previously computed residuals.
\end{enumerate}

\subsection{Case studies}
We compare the asymptotic variance of different estimators of the odds ratio in the simple case
of one binary risk variable for two data sets.
In this case, the odds ratio is a simple function of four counts given by Eq.~(\ref{or_quali}).
We focus on the simple random sampling without replacement and compare three estimators.
The first one is the Horvitz-Thompson estimator which
does not use the auxiliary variable and whose asymptotic variance is given by Eq.~(\ref{var_sans_inf}). The second estimator is the generalized regression estimator
which takes the auxiliary variable into account through a linear model fitting the linearized variable  against the auxiliary variable.
The third estimator is the $B$-spline calibration estimator with an asymptotic variance given by Eq.~(\ref{varasymbeta}).
In order to gain efficiency, the auxiliary variable is related to the linearized variable.
In the context of one binary factor, the linearized variable is given by Eq.~(\ref{lin1}) and takes four different values, which depend on the values of the variables $ X$ and $ Y$.
In order to be related to the linearized variable, the auxiliary variable is related to the product of the two variables
$X$ and $Y$, which is a strong property.
Moreover, because $u_{i,1}$, $X$, and $ Y$ are discrete, using auxiliary information
does not necessarily lead to an important gain in efficiency as the first health survey example will show.
The gain in efficiency however is significant in some cases.
In the second example using labor survey data, the gain in using the $B$-splines calibration estimator
compared to the Horvitz-Thompson estimator is significant because the auxiliary variable is related to
the variable $ Y$ but also to the factor $ X$; $X$ and $Y$ being related to one another, too.

\noindent \textit{Example from the California Health Interview Survey}

The data set comes from the Center for Health Policy Research at the University of California. It was extracted
from the adult survey data file of the California Health Interview Survey in 2009 and consists of 11074 adults.
The response dummy variable equals one if the person is currently insured; the binary factor equals one
if the person is currently a smoker. The auxiliary variable is age and we  consider people who are less than 60 years old.
The data are presented in  detail in Lumley (2010).

We compare the Horvitz-Thompson, the generalized regression, and the $B$-splines calibration estimators in terms of
asympotic variance. In order to calculate the $B$-splines functions, we use the SAS procedure {\em transreg} and take $K=15$ knots and $B$-splines of degree $m=3$.
The gain in using the generalized regression estimator compared to the Horvitz-Thompson estimator is only 0.01\%. It is 1.5\% when using
$B$-splines instead of the generalized regression.
When changing the number of knots and the degree of the $B$-spline functions, the results remain similar and the gain remains under 2\%.
In this example, there is no gain in using auxiliary information even with flexible $B$-splines, because the auxiliary variable
is not related enough to the linearized variable. The linearized variable
takes negative values for smokers without insurance and non smokers with insurance, positive values for smokers with insurance and non smokers without insurance. Age is not a good predictor for this variable,
because we expect to find sufficient people of any age in each of the four categories (smokers/non smokers $\times$ insurance/no insurance). Incorporating this auxiliary information brings no gain.

\noindent \textit{Example from the French Labor Survey}

We consider 14621
wage-earners under 50 years of age, from the French labour force survey. The initial data set consists of monthly wages in 2000 and 1999.
A dummy variable W00 equals one if the monthly wage in 2000
exceeds 1500 euros and zero otherwise. The same for W99 in 1999.
The population is divided in lower and upper education groups.
The value of the categorical factor DIP equals one for people with a university degree and zero otherwise.
W00 corresponds to the binary response variable $ Y$ while
the diploma variable DIP corresponds to the risk variable $ X$. The variable W99 is the auxiliary variable $Z$.

To compare the Horvitz-Thompson estimator with the generalized regression estimator and the $B$-splines calibration estimator, we calculate the gain in terms of
asympotic variance. We consider $K=15$ knots and the degree $m=3$.
The gain in using the generalized estimator compared to the Horvitz-Thompson estimator is now 20\%. It is 33\% when using $B$-splines.
The result is independent of the number of knots and, of the degree of $B$-spline functions. When the total number of knots varies
from 5 to 50 and the degree varies from 1 to 5, the gain is between 32\% and 34\%.
The nonlinear link between the linearized variable of a complex parameter with the
auxiliary variable explains the gain in using a nonparametric estimator compared to an estimator based on a linear model
(Goga and Ruiz-Gazen, 2013).
For the odds ratio with one binary factor, the linearized variable is discrete and the linear model does
not fit the data.

\section{Conclusion}

Estimating the variance of parameter estimators in a logistic regression is not straightforward especially if auxiliary
information is available. We applied the method of Goga and Ruiz-Gazen (2014) to the case of parameters defined through estimating equations.
The method relies on a linearization principle. The asymptotic variance of the estimator incorporates residuals of the model that we assume between
the linearized variable and the auxiliary variable. The gain in using auxiliary information is thus based on the fitting
quality of the model for the linearized variable. Because of the complexity of linearized variables, linear models 
that incorporate auxiliary information seldom fit linearized variables and we use nonparametric $B$-spline estimators. A particular case is post-stratification. Using the influence function defined by Eq.~(\ref{if}), we derive the asymptotic variance of the estimators together with confidence intervals for the odds ratio.\\

\noindent {\bf \large Acknowledgement:} we thank Beno\^it Riandey for drawing our attention to the odds ratio and one rewiever for his/her constructive comments.

\section*{Bibliography}

\begin{description}
\item Agarwal, G. G. and Studden, W. J. (1980),
\newblock{Asymptotic integrated mean square error using least squares and bias minimizing splines}.
\newblock{\em The Annals of Statistics}, {\em 8}: 1307-1325.

\item Agresti, A. (2002).
\newblock{\em Categorical Data Analysis}  (2nd edition).
\newblock{New York: John Wiley}.

\item Binder, D. A. (1983).
\newblock On the variance of asymptotically normal estimators from complex surveys.
\newblock{\em International Statistical Review}, {\em 51}: 279-292.

\item Deville, J.-C. (1999).
\newblock Variance estimation for complex statistics and estimators: linearization and residual techniques.
\newblock{\em Survey Methodology}, {\em 25}: 193-203.

\item Deville, J.-C. and S\"arndal, C.-E. (1992).
\newblock Calibration estimation in survey sampling.
\newblock{\em Journal of the American Statistical Association}, {\em 418}: 376-382.

\item Dierckx, P. (1993).
\newblock{\em Curves and Surfaces Fitting with Splines}.
\newblock{United Kingdom: Clarendon Press.}

\item Eideh, A. A. H. and Nathan, G. (2006).
\newblock The analysis of data from sample surveys under informative sampling.
\newblock{\em Acta et Commentationes Universitatis Tartuensis de Mathematica}, {\em 10}: 1-11.

\item Goga, C. (2005).
\newblock R\'eduction de la variance dans les sondages en pr\'esence d'information auxiliaire : une approche nonparam\'etrique par splines de r\'egression.
 \newblock{{\em The Canadian Journal of Statistics/Revue Canadienne de Statistique, 33}(2)}: 1-18.

\item Goga, C. and Ruiz-Gazen, A. (2014).
\newblock Efficient estimation of nonlinear finite population parameters using nonparametrics.
\newblock{\em Journal of the Royal Statistical Society series B}, \textbf{76}, 113-140. 

\item Heeringa, S. G., West, B. T., and Berglund, P. A. (2010).
\newblock{\em Applied Survey Data Analysis}. Chapman and Hall/CRC.

\item Horvitz, D .G. and Thompson, D. J. (1952).
\newblock A generalization of sampling without replacement from a finite universe.
\newblock{\em Journal of the American Statistical Association}, {\em 47}: 663-685.

\item Korn, E. L. and Graubard, B. I. (1999).
\newblock{\em Analysis of Health Survey}.
\newblock{New York: John Wiley}.

\item Lohr, S. L. (2010).
\newblock{\em Sampling: Design and Analysis}
\newblock{(2nd edition). Brooks/Cole, Cengage Learning}.

\item Lumley, T. (2010).
\newblock{\em Complex surveys: a guide to analysis using R}.
\newblock{New York: John Wiley}.

\item Rao, J. N. K., Yung, W., and Hidiroglou, M. A. (2002).
\newblock Estimating equations for the analysis of survey data using post-stratification information.
\newblock{\em Sankhya: The Indian Journal of Statistics}, {\em 64}: 364-378.

\item Ruppert, D., Wand, M. P., and Caroll, R.J. (2003).
\newblock{\em Semiparametric Regression}.
\newblock Cambridge Series in Statistical and Probabilistic Mathematics.
\newblock{New York: Cambridge University Press}.

\end{description}

\end{document}